\documentclass[12pt]{elsart}
\usepackage{graphics,bm}
\begin{document}
\begin{frontmatter}
\title{Chiral Symmetry in an Extended Constituent Quark Potential Model}
\author{Weizhen DENG},
\ead{dwz@pku.edu.cn}
\author{Yanrui LIU},
\author{Xiaolin CHEN},
\author{Dahai LU},
\author{Shi-Lin ZHU}
\address{%
Department of Physics, Peking University, BEIJING 100871, CHINA}

\date{\today}

\begin{abstract}
The chiral symmetry is applied to an extended constituent quark
potential model. With random phase approximation (RPA), a small
component effect is added to the constituent quark model. To obtain
the pseudoscalar $\pi$ meson as a Goldstone boson, the quark effective
potentials are modified in the model to account for the dynamical
breaking of chiral symmetry.  Also the vector $\rho$ meson is
calculated and the KSRF relation about $\pi$ and $\rho$ meson decay
constants is derived in the model.
\end{abstract}

\begin{keyword}
chiral symmetry \sep quark potential model \sep random phase approximation
\PACS 12.39.Pn \sep 14.40.Aq
\end{keyword}
\end{frontmatter}

\section{Introduction}

Due to the complication of a non-Abelian $SU(3)$ gauge theory, quantum
chromodynamics (QCD), which describes the strong interaction, has many
non-perturbative features such as the dynamical breaking of chiral
symmetry and quark confinement. In the study of hadron structure at
low energy scale, two kinds of models of QCD are often used, each
incorporating some important QCD features.  The constituent quark
potential model which incorporates the QCD quark confinement has been
impressively successful in hadron spectroscopy and decays
\cite{Isgur85}, except for the pseudo-scalar $\pi$ meson which is a
Goldstone boson and has very low mass.  It was shown recently that the
RPA $y$-component must be considered in the treatment of pions
\cite{Cotanch00}.  On the other hand, the Nambu-Jona-Lasinio (NJL)
model \cite{Nambu61} and several of its extensions
\cite{Bernard84,Klimt90,Langfeld96} which incorporate the chiral
symmetry can describe the $\pi$ meson very well as a massless
Goldstone boson of dynamical breaking of chiral symmetry.  However,
the NJL model lacks the important QCD property of quark confinement.

In Ref.~\cite{DWZ02}, we proposed an extension to the constituent
quark model to comprise the RPA $y$-component quark excitations.
As in NJL model, we start from an effective quark Hamiltonian.
For meson structure, this leads to a Hamiltonian with two channels
--- the ordinary valence quark $x$-channel and a new $y$-channel,
which is an extension of the quark potential model.  This extended
quark potential model has a new coupling potential connecting the
ordinary valence quark $x$-channel and the small component
$y$-channel. The coupling potential can be related to the ordinary
valence quark potential in the model.

In this paper, we implement the chiral symmetry in the extended
constituent quark potential model.  First we introduce the extended
quark potential model.  In Sec.~3, we analyze $\pi$ and $\rho$ mesons
using this framework. The restriction from chiral symmetry is
considered. With the meson wave functions containing $y$-channel quark
excitations, KSRF relation \cite{KS1966,RF1966} about $\pi$ and $\rho$
meson decay constants is derived. Finally, we summarize our
discussion.

\section{Extending Quark Potential Model with RPA}

QCD perturbation theory is very successful when applied in the
high energy processes. In the low energy regime, Hamiltonian
approach can provide an alternative method to understand the
non-perturbative features of QCD, such as the quark confinement.
Many efforts were made to obtain an effective Hamiltonian from the
exact QCD Lagrangian \cite{Wilson94,Cotanch99,Cotanch02}.  Here
we start from the effective quark hamiltonian, assuming the quark
fields have been separated approximately from the gluon fields
\begin{eqnarray}
\label{H}
H &=& H_2 + H_4, \\
H_2 &=& \int d^3x \Psi^\dagger(\bm{x}) (-i \bm{\alpha} \cdot \bm{\nabla}
         + \beta m_0) \Psi(\bm{x}), \\
H_4 &=& \frac12 \sum\limits_{\epsilon} \int d^3x d^3y
K_\epsilon(\bm{x} - \bm{y})
         \bar\Psi(\bm{x}) \Gamma_\epsilon \Psi(\bm{x})
         \bar\Psi(\bm{y}) \Gamma_\epsilon \Psi(\bm{y}),
\end{eqnarray}
where $H_2$ is the free current quark Hamiltonian and $m_0$ is the
current quark mass. $H_4$ includes all effective quark
interactions. $K_\alpha (\bm{x})$ is the kernel function for
coupling vertex $\Gamma_\alpha$.

Apart from a constant, the Hamiltonian (\ref{H}) can be written in
normal order as
\begin{equation}
\label{HP}
H = H_0 + :H_4:
\end{equation}
where $H_0$ represents the free particle energy of constituent quark
and its general form is
\begin{equation}
H_0 = \int d^3x \Psi^\dagger(\bm{x}) \left[ -i A(- \bm{\nabla}^2)
      \bm{\alpha} \cdot \bm{\nabla} + \beta B(- \bm{\nabla}^2) \right]
      \Psi(\bm{x}),
\end{equation}
where the dependence on the three-vector momentum arises from the
instantaneous approximation of quark interactions.

The two functions $A(\bm{k}^2)$ and $B(\bm{k}^2)$ should be
obtained from a self-consistent mean field calculation.  However
in a constituent quark potential model, the quark interactions
contain the linear quark confinement which has severe infrared
divergence in momentum space. On the other hand, the constituent
potential quark models always assume that the constituent quarks
have fixed masses. As an extension of the constituent potential
quark model, here we assume that the $A$ and $B$ functions can be
approximated as constants. So we treat the constituent quark mass
as a parameter, $m_c=B$ (with $A=1$). The free particle
Hamiltonian becomes
\begin{equation}
\label{H00}
H_0 = \int d^3x \Psi^\dagger(\bm{x}) \left[ -i
      \bm{\alpha} \cdot \bm{\nabla} + \beta m_c \right]
      \Psi(\bm{x}).
\end{equation}

In this paper, we will only consider meson structures.  In na\"ive
quark model, meson fields are approximately expressed in terms of
local quark fields as
\begin{equation}
\phi_M(x) = \alpha_M \bar\Psi(x) \Gamma_M \Psi(x),
\end{equation}
where $\Gamma_M=i\gamma_5 \tau_i$ for iso-vector pseudo-scalar $\pi$
mesons or $\Gamma_M=\gamma\cdot\epsilon_i \tau_j$ for iso-vector
vector $\rho$ mesons. According to $1/N$ expansion, the above meson
structures are exact in the limit $N_c \rightarrow \infty$
\cite{Hooft1974,Witten79}.

In the non-relativistic limit, as quark momentum $\bm{k} \ll k^0$, one
has\cite{DWZ01}
\begin{eqnarray}
\label{Meson-F1}
\phi_M(\bm{x}) &=& \alpha_M \sum_{s_1s_2} \int \frac{d^3k_1}{(2\pi)^3}
\frac{d^3k_2}{(2\pi)^3} \frac1{2k^0_1} \frac1{2k^0_2}
e^{-i(\bm{k}_1+\bm{k}_2)\cdot \bm{x}} \nonumber \\
&\times& \left[ \bar{u}(-\bm{k}_1, s_1) \Gamma_M v(-\bm{k}_2, s_2)
b^\dagger(-\bm{k}_1, s_1) d^\dagger(-\bm{k}_2, s_2) \right.
\nonumber \\
&& + \left. \bar{v}(\bm{k}_1, s_1) \Gamma_M u(\bm{k}_2, s_2)
d(\bm{k}_1, s_1) b(\bm{k}_2, s_2) \right],
\end{eqnarray}
for $\pi$ and $\rho$ mesons. On the other hand, the field operators of
$\pi$ and $\rho$ can also be expressed in the $Q$ and $Q^\dagger$ of
meson creation and annihilation operators
\begin{equation}
\label{Meson-F2}
\phi_M(x) = \int \frac{d^3k}{(2\pi)^3} \frac1{2k^0}
\left[ Q_M(\bm{k}) e^{-i k\cdot x} + Q_M^\dagger(\bm{k}) e^{i k\cdot
    x} \right].
\end{equation}
If the mesons are composed of creation of valence quark pairs as in
the potential models, one deduces from the equivalence of
eqs. (\ref{Meson-F1}) and (\ref{Meson-F2})
\begin{eqnarray}
Q^+_M(\bm{k}) &=& \alpha_M \sum_{s_1s_2} \int \frac{d^3k_1}{(2\pi)^3}
\frac{d^3k_2}{(2\pi)^3} \frac1{2k_1^0} \frac1{2k_2^0} (2\pi)^3
\delta^3(\bm{k}_1 + \bm{k}_2 - \bm{k}) \nonumber \\
&&\times \bar{u}(\bm{k}_1, s_1) \Gamma_M v(\bm{k}_2, s_2)
b^\dagger(\bm{k}_1, s_1) d^\dagger(\bm{k}_2, s_2).
\end{eqnarray}
However, generally one can only obtain
\begin{eqnarray}
Q^+_M(\bm{k}) &=& \alpha_M \sum_{s_1s_2} \int \frac{d^3k_1}{(2\pi)^3}
\frac{d^3k_2}{(2\pi)^3} \frac1{2k_1^0} \frac1{2k_2^0} (2\pi)^3
\delta^3(\bm{k}_1 + \bm{k}_2 - \bm{k}) \nonumber \\
&\times& \left[ X_M \bar{u}(\bm{k}_1, s_1) \Gamma_M v(\bm{k}_2, s_2)
b^\dagger(\bm{k}_1, s_1) d^\dagger(\bm{k}_2, s_2) \right .
\nonumber \\
&&+ \left. Y_M \bar{v}(-\bm{k}_1, s_1) \Gamma_M u(-\bm{k}_2, s_2)
d(-\bm{k}_1, s_1) b(-\bm{k}_2, s_2) \right].
\end{eqnarray}
This is just the RPA excitation operator --- the meson state is a
superposition of creation and annihilation of $q\bar{q}$ pairs on
the vacuum. The second part, i.e. the $y$-component is simply
discarded in the ordinary quark potential model as a small
component.

Thus, as an extension to the potential model, we take the mesons as
excitation modes of the vacuum of the RPA type:
\begin{equation}
\left| Q \right\rangle = Q^\dagger \left| 0 \right\rangle.
\end{equation}
For a meson in the rest frame, after we take account of the quark
interactions, $Q^\dagger$ excitation operators are
\begin{eqnarray}
\label{Qop}
Q^\dagger & = & \sum_{s_1 s_2} \int \frac{d^3k}{(2\pi )^3} \frac 1{2k^0}
 \left[ x(\bm{k},s_1,s_2)
 b^\dagger(\bm{k} s_1) d^\dagger(-\bm{k} s_2) \right. \nonumber \\
 &  & \left. + y(\bm{k},s_1,s_2)
 \widetilde{d}(-\bm{k} s_1) \widetilde{b}(\bm{k} s_2) \right],
\end{eqnarray}
where the time reversals are defined as
\begin{eqnarray}
\widetilde{b}(\bm{k} s) &\equiv& (-1)^{1/2-s} b(\bm{k} -s), \\
\widetilde{d}(\bm{k} s) &\equiv& (-1)^{1/2-s} d(\bm{k} -s),
\end{eqnarray}
and $x$ and $y$ are the RPA amplitudes determined from the well-known
RPA equation of motion\cite{Ring80}
\begin{equation}
\label{EOM}
\left\langle 0 \right| \left[ \delta Q,[H,Q^\dagger] \right]
\left| 0 \right\rangle
 = E \left\langle 0 \right| [\delta Q,Q^\dagger]
\left| 0 \right\rangle.
\end{equation}
We obtain the equations for RPA amplitudes
\begin{eqnarray}
\label{XRPA-M}
&&\left(E - 2 k^0 \right) x(\bm{k},s_1,s_2) \nonumber \\
&=& \sum_{s_3s_4} \int \frac{d^3k^{\prime}}{(2\pi)^3}
\left[ \langle \bm{k},s_1,s_2 \mid U \mid \bm{k}',s_3,s_4 \rangle
             x(\bm{k}',s_3,s_4) \right. \nonumber \\
& & \left. +\langle \bm{k},s_1,s_2 \mid V \mid \bm{k}',s_3,s_4 \rangle
             y(\bm{k}',s_3,s_4) \right],
\\
\label{YRPA-M}
&&\left(E + 2 k^0 \right) y(\bm{k},s_1,s_2) \nonumber \\
&=& -\sum_{s_3s_4} \int \frac{d^3k^{\prime}}{(2\pi)^3}
\left[ \langle \bm{k},s_1,s_2 \mid V \mid \bm{k}',s_3,s_4 \rangle
             x(\bm{k}',s_3,s_4) \right. \nonumber \\
& & \left. +\langle \bm{k},s_1,s_2 \mid U \mid \bm{k}',s_3,s_4 \rangle
             y(\bm{k}',s_3,s_4) \right].
\end{eqnarray}
Here we have introduced two effective potentials $U$ and $V$.  Their matrix
elements are
\begin{eqnarray}
\label{U}
\langle \bm{k},s_1,s_2 \mid U \mid \bm{k}',s_3,s_4 \rangle
& \equiv & -\frac1{2k^0} \frac1{2k^{\prime 0}}
     \sum_{\epsilon} K_{\epsilon} (\bm{k}-\bm{k}^\prime)
     \nonumber \\
&  & \bar{u}(\bm{k}s_1) \Gamma_\epsilon u(\bm{k}'s_3)
     \bar{v}(-\bm{k}'s_4) \Gamma_\epsilon v(-\bm{k}s_2), \\
\label{V}
\langle \bm{k},s_1,s_2 \mid V \mid \bm{k}',s_3,s_4 \rangle
& \equiv & \frac1{2k^0} \frac1{2k^{\prime 0}}
     \sum_{\epsilon} K_{\epsilon} (\bm{k}-\bm{k}^\prime)
     \nonumber \\
&  & \bar{u}(\bm{k}s_1) \Gamma_\epsilon \widetilde{v}(-\bm{k}'s_3)
     \widetilde{\bar{u}}(\bm{k}'s_4) \Gamma_\epsilon v(-\bm{k}s_2).
\end{eqnarray}

Thus, with the RPA approximation, the quark potential model is
extended to a coupling system with two channels. The first part,
i.e. x-amplitude $x(\bm{k},s_1,s_2)$ which we will call the x-channel
wave function, is the large component which is the sole valence quark
contribution in the ordinary quark potential model. The second part,
i.e. the y-amplitude $y(\bm{k},s_1,s_2)$ which we will call the
y-channel wave function, is the small component which is discarded in
the ordinary quark potential model. The potential $U$ interacts only
within each individual channel while the new coupling potential $V$
couples the two channels together. If the coupling between the two
channels $V$ is small, the system decouples. The extended potential
model reduces to the ordinary potential model and the potential $U$ is
just the ordinary quark potential. Here we will show that the coupling
potential $V$ is important due to chiral symmetry .

One can use the Dirac kets and bras to simplify
notations. $x(\bm{k},s_1,s_2)$ and $y(\bm{k},s_1,s_2)$ are
represented by two kets $|x\rangle$ and $|y\rangle$ respectively,
\begin{eqnarray}
x(\bm{k},s_1,s_2) &=& \langle\bm{k},s_1,s_2 \mid x \rangle, \\
y(\bm{k},s_1,s_2) &=& \langle\bm{k},s_1,s_2 \mid y \rangle.
\end{eqnarray}
Eqs. (\ref{XRPA-M}) and (\ref{YRPA-M}) can be written concisely as
\begin{eqnarray}
\left(E - 2 k^0 \right) |x\rangle &=& U |x\rangle + V |y\rangle ,
\label{EQM-XRPA}\\
\left(E + 2 k^0 \right) |y\rangle &=& - V |x\rangle - U |y\rangle ,
\label{EQM-YRPA}
\end{eqnarray}
or the familiar matrix form
\begin{equation}
\label{EQM-RPA}
\left[\begin{array}{cc} 2k^0 + U & V \\ -V & -2k^0 - U \end{array}\right]
\left[\begin{array}{c} |x\rangle \\ |y\rangle \end{array}\right]
= E
\left[\begin{array}{c} |x\rangle \\ |y\rangle \end{array}\right],
\end{equation}
where $k^0 = \sqrt{m_c^2 + \bm{k}^2}$.

By definitions (\ref{U}) and (\ref{V}), the potentials $U$ and $V$
will be evaluated from the same kernels in the quark interaction
$H_4$, thus they can be related to each other. In the
non-relativistic limit $\bm{k}, \bm{k}' \to 0$, we have
\begin{eqnarray}
\label{U-Meson}
U &=& S - V - A \bm{\sigma}_1 \cdot \bm{\sigma}_2
- T \bm{\sigma}_1 \cdot \bm{\sigma}_2, \\
V &=& P - V \bm{\sigma}_1 \cdot \bm{\sigma}_2 - A
- T \bm{\sigma}_1 \cdot \bm{\sigma}_2,
\end{eqnarray}
where on the right hand side of equations, the kernels are divided into
five types as in ref.~\cite{Gromes77}, i.e., scalar ($S$),
pseudo-scalar ($P$), vector ($V$), axial-vector ($A$), and tensor ($T$);
$\bm{\sigma}$'s are the pauli spin matrices.

To normalize the model wave functions, we apply the standard boson
commutation relation
\begin{equation}
[Q(P), Q^\dagger(P')] = (2\pi)^3 2P^0 \delta^3(\bm{P}-\bm{P}')
\end{equation}
to rest mesons. We obtain
\begin{equation}
\langle Q \mid Q \rangle = (2\pi)^3 2E \delta^3(0).
\end{equation}
Then we find the normalization relation of RPA type
\begin{equation}
\label{Normalization}
2E = \sum_{s_1, s_2} \int\frac{d^3k}{(2\pi)^3} \left\{
|x(\bm{k}, s_1, s_2)|^2 - |y(\bm{k}, s_1, s_2)|^2 \right\}.
\end{equation}
Again with Dirac kets and bras, its form is
\begin{equation}
2E = \langle x \mid x \rangle - \langle y \mid y \rangle.
\end{equation}

With the normalized meson wave functions, one can calculate the
static properties of mesons.  The $\pi$ weak decay constant is
defined as
\begin{equation}
if_\pi q^\mu \delta_{ij} = \langle 0 \mid
\bar\Psi(0) \gamma^\mu \gamma_5 \frac12 \tau_i \Psi(0) \mid
\pi_j(q) \rangle.
\end{equation}
For pions at rest, the above equation becomes
\begin{equation}
if_\pi m_\pi \delta_{ij} = \langle 0 \mid
\bar\Psi(0) \gamma^0 \gamma_5 \frac12 \tau_i \Psi(0) \mid
Q_{\pi j} \rangle.
\end{equation}
In the non-relativistic limit, we obtain
\begin{equation}
i f_\pi m_\pi = \langle \bm{r}=0, S=0,M_S=0 \mid x \rangle
+ \langle \bm{r}=0, S=0,M_S=0 \mid y \rangle,
\end{equation}
where $\langle \bm{r}, S=0,M_S=0 \mid x \rangle$ and $\langle
\bm{r}, S=0,M_S=0 \mid y \rangle$ are the coordinate wave
functions in x- and y- channel of the $\pi$ meson with its total
spin $S=0,M_S=0$.  The $\rho$ electro-magnetic decay constant is
defined as
\begin{equation}
\frac{m_\rho^2}{f_\rho} \epsilon_k^\mu \delta_{ij} = \langle 0 \mid
\bar\Psi(0) \gamma^\mu \frac12 \tau_i \Psi(0) \mid
\rho_{jk}(q) \rangle,
\end{equation}
where $i, j$ are the isospin indices, and $k$ is the $\rho$ meson
polarization index. For static $\rho$ mesons
\begin{equation}
\frac{m_\rho^2}{f_\rho} \delta_{ij} = \langle 0 \mid
\bar\Psi(0) \gamma^k \frac12 \tau_i \Psi(0) \mid
Q_{\rho jk} \rangle.
\end{equation}
In the non-relativistic limit, one obtains
\begin{equation}
\frac{m_\rho^2}{f_\rho} =
\langle \bm{r}=0, S=1,M_S \mid x \rangle
+ \langle \bm{r}=0, S=1,M_S \mid y \rangle.
\nonumber \\
\end{equation}

\section{$\pi$ and $\rho$ meson Properties}

Following the quark potential model \cite{Isgur85},
we choose the effective quark interaction $H_4$ as a vector
interaction plus a scalar confinement interaction
\begin{eqnarray}
\label{Cal-H4}
H_4 &=&
\frac12 \int d^3x d^3y K_s(\bm{x} - \bm{y})
\bar\Psi(\bm{x}) \Psi(\bm{x})
\bar\Psi(\bm{y}) \Psi(\bm{y}) \nonumber\\
&+&\frac12 \int d^3x d^3y K_v(\bm{x} - \bm{y})
\bar\Psi(\bm{x}) \gamma^\mu \frac{\bm{\lambda}^c}{2} \Psi(\bm{x})
\bar\Psi(\bm{y}) \gamma_\mu \frac{\bm{\lambda}^c}{2} \Psi(\bm{y})
.
\end{eqnarray}
According to lattice calculation, the scalar confinement kernel is
linear in $r$
\begin{equation}
K_s(r) = br.
\end{equation}
The vector kernel is taken from the one-gluon exchange plus a constant
\begin{equation}
K_v(r) = c + \frac{\alpha_s(r)}{r},
\end{equation}
where $\alpha_s(r)$ is the QCD running coupling constant.

In our analysis, we use the simple harmonic oscillator wave
functions for pions and rho mesons as a first approximation. In
this way the underlying physics is emphasized.
\begin{eqnarray}
\label{Test-X-WVF}
x(\bm{k}, s_1, s_2) &=& X \Psi_{000}(k)
\langle \frac12 s_1 \frac12 s_2 \mid S M_S \rangle, \\
\label{Test-Y-WVF}
y(\bm{k}, s_1, s_2) &=& Y \Psi_{000}(k)
\langle \frac12 s_1 \frac12 s_2 \mid S M_S \rangle,
\end{eqnarray}
where $S$, $M_S$ are the spin quantum numbers of the meson, and
$\Psi_{000}(k)$ is the ground state wave function of harmonic
oscillator in momentum space
\begin{equation}
\Psi_{000}(k) = \frac{1}{(\sqrt{\pi}\beta)^{3/2}} e^{- k^2 / 2\beta^2}.
\end{equation}

In the non-relativistic limit, the two potentials $U$ and $V$ are
\begin{eqnarray}
\label{U-pot-simp}
U &=&  br +
\left(c + \frac{\alpha_s}{r}\right) \bm{F}_1 \cdot \bm{F}_2, \\
\label{V-pot-simp}
V &=&  c \bm{\sigma}_1 \cdot \bm{\sigma}_2
\bm{F}_1 \cdot \bm{F}_2 ,
\end{eqnarray}
where
\begin{equation}
\bm{F_i} =
\left\{ \begin{array}{ll}
\frac{\bm{\lambda}_i}{2} & \qquad\mbox{for quarks}, \\
\frac{\bm{\lambda}^c_i}{2} = -\frac{\bm{\lambda}^*_i}{2} & \qquad
\mbox{for antiquarks}. \end{array}\right.
\end{equation}
Here, one must be careful with the one-gluon exchange interaction
$\frac{\alpha_s}{r}$. It arises from the covariant form
$\frac{4\pi\alpha_s}{Q^2}$ in momentum space, where $Q=k-k'$ is
the 4-momentum of the exchanged gluon. For potential $U$,
$k=(k^0,\bm{k})$ and $k'=(k^{\prime0}, \bm{k}')$ are the 4-momenta
of incoming and outgoing quarks respectively. Thus in
non-relativistic limit, $\frac{4\pi}{Q^2} \approx
\frac{4\pi}{\bm{Q}^2}$ which is a Coulomb potential in coordinate
space. However for coupling potential $V$, $k=(k^0, \bm{k})$ and
$-k'=(k^{\prime0}, -\bm{k}')$ are the 4-momenta of $q\bar{q}$
quark pair which were created or annihilated on the vacuum.  So in
non-relativistic limit, $\frac{4\pi}{Q^2} \approx
\frac{4\pi}{(2m_c)^2} \approx 0$. The term $\frac{\alpha_s}{r}$ is
suppressed in $V$.

It is well known in NJL model, the chiral symmetry can be broken
dynamically with a nonzero vacuum condensate coming from quark
interaction.  The current quarks gain dynamical masses and become
constituent quarks which are the valence quarks in potential
model. The Goldstone boson pions are still massless.  To account for
this dynamical chiral symmetry breaking in our model, the coupling
potential $V$ must be large.  This is why we introduce a constant term
into the vector kernel.  In this way, our model differs from the
ordinary potential model (see ref.~\cite{Isgur85}) which had the
constant term in the scalar interaction.  In an ordinary potential
model, this change will cause no difference (the potential $U$ is not
changed). But in this extended potential model, the coupling potential
$V$ can be made strong enough to give $\pi$ mesons very low masses.

\begin{figure}
\includegraphics{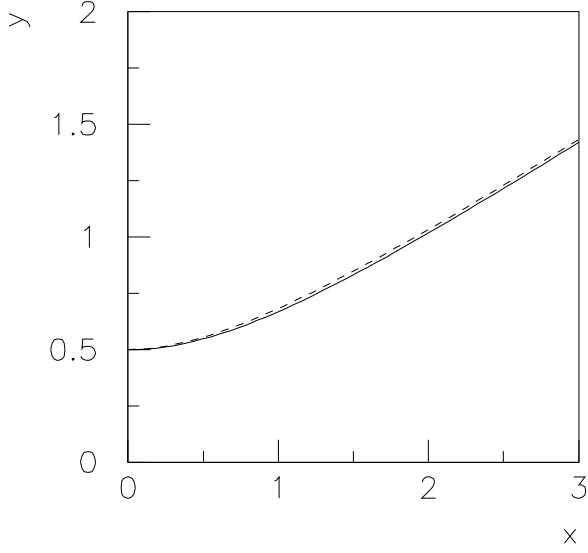}
\caption{\label{eps-1}Comparison of the integral $\int_0^\infty
t^2 dt \sqrt{t^2 + x^2} \exp(-t^2)$ (solid line) in the average
quark kinetic energy with $\frac12\sqrt{1+\frac{\pi}4 x^2}$
(dashed line) used in Eq. (\ref{kk}).}
\end{figure}

Inserting the simple model wave functions (\ref{Test-X-WVF}) and
(\ref{Test-Y-WVF}) into Eq. (\ref{EQM-RPA}) with the potentials
given by (\ref{U-pot-simp}) and (\ref{V-pot-simp}), we obtain an
eigen equation of a $2\times2$ matrix
\begin{equation}
\label{EQM-RPA-simp}
\left[\begin{array}{cc} A  & B \\ -B & -A \end{array}\right]
\left[\begin{array}{c} X \\ Y \end{array}\right]
= E
\left[\begin{array}{c} X \\ Y \end{array}\right],
\end{equation}
where
\begin{eqnarray}
A &=& 2 \langle \sqrt{\bm{k}^2 + m_c^2} \rangle +
\frac{2b}{\sqrt{\pi}\beta} - \frac{4c}{3}
- \frac43 \left\langle \frac{\alpha_s(r)}{r}\right\rangle, \\
B &=& -\frac{4c}{3} \langle \bm{\sigma}_1
\cdot \bm{\sigma}_2 \rangle .
\end{eqnarray}
The energy $E$ can be easily obtained
\begin{equation}
E = \sqrt{A^2-B^2}.
\end{equation}
The wave function parameter $\beta$ is determined by the minimum of the
energy
\begin{equation}
\frac{\partial E}{\partial \beta} = 0.
\end{equation}
Since $B$ is a constant, we get
\begin{equation}
\frac{\partial A}{\partial \beta} = 0.
\end{equation}
We approximate the average quark kinetic energy by the following
expression.
\begin{equation}\label{kk}
\langle K \rangle_{m_c} = \langle \sqrt{\bm{k}^2 + m_c^2} \rangle \approx
\sqrt{\frac4{\pi}\beta^2 + m_c^2}.
\end{equation}
Numerically the deviation is very small and less than 5\% (See
Fig. \ref{eps-1}).

The current quark mass $m_0$ can be obtained from the partial conservation
of axial-vector current (PCAC). Consider the axial-vector current
\begin{equation}
A_{ud}^\mu (x) = \bar{u} (x) \gamma^\mu \gamma_5 d(x).
\end{equation}
According to PCAC, its divergence is
\begin{equation}
\label{PCAC}
\partial_\mu A_{ud}^\mu(x) = (m_u+m_d) \bar{u}(x) i\gamma_5 d(x),
\end{equation}
where $m_u$ and $m_d$ are current masses of up and down quarks.
Let us calculate the matrix element $\langle 0 \mid
\partial_\mu A_{ud}^\mu(x) \mid \pi^+ \rangle$ for $\pi^+$ meson with
momentum $q_\pi$.
One obtains
\begin{equation}
\langle 0 \mid \partial_\mu A_{ud}^\mu(x) \mid \pi^+, q_\pi \rangle
= -i q_\pi^\mu \langle 0 \mid A_{ud}^\mu(x) \mid \pi^+, q_\pi \rangle.
\end{equation}
>From right hand side of PCAC relation (\ref{PCAC}), one also has
\begin{equation}
\langle 0 \mid \partial_\mu A_{ud}^\mu(x) \mid \pi^+ \rangle
= (m_u+m_d) \langle 0 \mid \bar{u}(x) i\gamma_5 d(x) \mid \pi^+ \rangle.
\end{equation}
Both of the matrix elements can be easily calculated using the
model wave functions for a static $\pi^+$ meson. We obtain
\begin{equation}
\label{PCAC-m0}
2m_0 \equiv m_u+m_d = \frac{X_\pi+Y_\pi}{X_\pi-Y_\pi} m_\pi.
\end{equation}

Before the numerical calculation, one can make some qualitative
analysis of the properties of $\pi$ and $\rho$ mesons in the
model. Let
\begin{eqnarray}
u&=&X+Y,\\
v&=&X-Y.
\end{eqnarray}
The matrix equation (\ref{EQM-RPA-simp}) can be written for $u$, $v$
\begin{eqnarray}
(A-B)v &=& E u, \\
(A+B)u &=& E v.
\end{eqnarray}
Also the normalization relation (\ref{Normalization}) becomes
\begin{equation}
X^2 - Y^2 = u v = 2E.
\end{equation}

First, let us consider the $\pi$ meson which is a Goldstone boson.
We have
\begin{eqnarray}
\label{eqn-pi-1}
\left (2A_0 - \frac{16c}{3} \right)
v_\pi &=& m_\pi u_\pi, \\
\label{eqn-pi-2}
\left (2A_0 + \frac{8c}{3} \right)
u_\pi &=& m_\pi v_\pi,
\end{eqnarray}
where
\begin{equation}
A_0 = \sqrt{\frac{4}{\pi} \beta^2 + m_c^2} + \frac{b}{\sqrt{\pi}\beta}
- \frac23 \left\langle\frac{\alpha_s(r)}{r}\right\rangle.
\end{equation}
If the current quark mass $m_0=0$, chiral symmetry is a strict QCD
symmetry.  From the Goldstone theorem, $\pi$ meson should be
massless. As an effective theory of QCD, the $\pi$ meson mass in
our model should also be zero. Assuming $m_c$ is the constituent
quark mass in the limit of $m_0=0$, we have
\begin{equation}
\label{eqn-K}
2\left. A_0 \right|_{m_c} - \frac{16c}{3} = 0.
\end{equation}
In the real world, current quarks have small nonzero masses. So
does the $\pi$ meson. As the average mass of current quarks (u and
d) slightly changes to $m_0\ne 0$, the constituent quark mass will
also change a little bit to $m_c+\Delta m_c$ with $\Delta m_c \ll
m_c$. So we have
\begin{equation}
\left. A_0 \right|_{m_c+\Delta m_c} \approx \left. A_0 \right|_{m_c} \equiv
A_0
\end{equation}
and
\begin{equation}
\left. A_0 \right|_{m_c+\Delta m_c} - \left. A_0 \right|_{m_c}
\approx \frac{\partial\langle K \rangle_{m_c}}{\partial m_c} \cdot \Delta m_c
= \frac{m_c}{\langle K \rangle_{m_c}} \cdot \Delta m_c.
\end{equation}
Eqs. (\ref{eqn-pi-1}) and (\ref{eqn-pi-2}) become
\begin{eqnarray}
2\frac{\Delta m_c m_c}{\langle K \rangle_{m_c}} v_\pi &=& m_\pi u_\pi, \\
3A_0 u_\pi &=& m_\pi v_\pi.
\end{eqnarray}
We obtain
\begin{eqnarray}
m_\pi &=& \sqrt{\frac{6\Delta m_c m_c A_0}{\langle K \rangle_{m_c}}},\\
\label{uv_pi}
\frac{u_\pi}{v_\pi} &=& \frac{m_\pi}{3 A_0}.
\end{eqnarray}
With the normalization $2m_\pi = u_\pi v_\pi$, we obtain
\begin{eqnarray}
u_\pi &=& \sqrt{\frac23} \frac{m_\pi}{\sqrt{A_0}},\\
v_\pi &=& \sqrt{6 A_0}.
\end{eqnarray}
>From PCAC (eq.~(\ref{PCAC-m0})), the mass of current quark is related
to $\Delta m_c$ as
\begin{equation}
m_0 = \frac{m_c}{\langle K \rangle_{m_c}} \Delta m_c.
\end{equation}
Next we can calculate the weak decay constant of $\pi$ meson to obtain
\begin{equation}
\label{f_pi_cal}
f_\pi = \sqrt{\frac{2N_c}{3 A_0}
\left(\frac{\beta}{\sqrt{\pi}} \right)^3}.
\end{equation}
Eq. (\ref{uv_pi}) means $u_\pi/v_\pi \approx 0$ and $Y_\pi \approx
- X_\pi$, which shows the importance of the contribution to the
$\pi$ meson from the small component of y-channel .

For the $\rho$ meson, with the help of Eq. (\ref{eqn-K}), the
equations are
\begin{eqnarray}
2A_0 v_\rho &=& m_\rho u_\rho, \\
A_0 u_\rho &=& m_\rho v_\rho.
\end{eqnarray}
One simply obtains
\begin{eqnarray}
m_\rho &=& \sqrt2 A_0, \\
\label{uv_rho}
\frac{u_\rho}{v_\rho} &=& \sqrt2.
\end{eqnarray}
With the normalization $2m_\rho = u_\rho v_\rho$,
\begin{eqnarray}
u_\rho &=& 2\sqrt{A_0} \\
v_\rho &=& \sqrt{2 A_0}.
\end{eqnarray}
Next the electromagnetic decay constant of $\rho$ meson can be
calculated as
\begin{equation}
\label{f_rho_cal}
f_\rho^{-1} = \frac{1}{m_\rho}
 \sqrt{\frac{2N_c}{A_0}\left(\frac{\beta}{\sqrt{\pi}} \right)^3}.
\end{equation}

A relation among masses of $\pi$, $\rho$ mesons and current quark mass
$m_0$ can be easily obtained
\begin{equation}
m_\pi^2 = 6m_0 A_0 = 3\sqrt2 m_0 m_\rho.
\end{equation}
>From the experimental $\rho$ meson mass $m_\rho=770$ MeV, one can
estimate that $A_0 \approx m_\rho/\sqrt2 \approx 540$ MeV. Then we
can estimate the current quark mass to be $m_0 \approx 6$ MeV from
the physical pion mass $m_\pi=140$ MeV.

Another relation about $\pi$ and $\rho$ decay constants can also
be easily obtained from Eqs. (\ref{f_pi_cal}) and
(\ref{f_rho_cal}),
\begin{equation}
f_\pi f_\rho = \frac1{\sqrt3} m_\rho.
\end{equation}
The KSRF relation \cite{KS1966,RF1966} gives
\begin{equation}
f_\pi f_\rho = \frac1{\sqrt2} m_\rho,
\end{equation}
while the experiment value is $m_\rho / (f_\pi f_\rho) = 1.62$. In our
approach $Y_\rho/X_\rho=1/(3+\sqrt2)=0.17$ from Eq.
(\ref{uv_rho}). This means the contribution to the rho meson from
y-channel is rather small. If the small component is completely
neglected, i.e., $Y_\rho=0$, we may obtain $m_\rho = \frac32 A_0$ and
$X_\rho = \sqrt{2 m_\rho}$. Then
\begin{equation}
f_\rho^{-1} = \frac{1}{m_\rho}
\sqrt{\frac{4N_C}{3A_0} \left(\frac{\beta}{\sqrt{\pi}}\right)^3},
\end{equation}
thus
\begin{equation}
f_\pi f_\rho = \frac1{\sqrt2} m_\rho,
\end{equation}
which is exactly the KSRF relation. The experimental value is
between our result and KSRF relation.

In numerical calculation we parametrize the running coupling constant as
\begin{eqnarray}
\alpha_s(\bm{Q}^2) &=& \sum_{k} \alpha_k \exp(-\bm{Q}^2/4\gamma_k^2)
\nonumber\\
&=& 0.25 \exp(-\bm{Q}^2) + 0.15 \exp(-\bm{Q}^2/10) + 0.20
\exp(-\bm{Q}^2/1000)
\end{eqnarray}
as in Ref.~\cite{Isgur85} (with $Q$ in GeV).  Also the
constituent quark mass $m_c$ is fixed to be $m_c=220$ MeV.  The
results are listed in Table \ref{table-1}. First (Set 1), we
choose the confinement parameter $b=0.18$GeV$^2$ according to
lattice calculation. The $\rho$ meson mass is somehow larger. Next
(Set 2), we adjust the confinement parameter slightly to $b =
0.15$GeV$^2$.
\begin{table}
\caption{\label{table-1}Numerical results for $\pi$ and $\rho$ mesons.
Experimetal data are taken from ref.~\cite{PDG2000}.}
\begin{tabular}{lcccccc}
\hline
\hline
&$b$&$c$&$\beta$&$m_0$&$m_\pi$&$m_\rho$ \\
&(GeV$^2$)&(MeV)&(MeV)&(MeV)&(MeV)&(MeV)\\
\hline
Exp. &&&&&$139$&$770$\\
\hline
Set 1 &$0.18$&$220$&$391$&$6.5$&$152$&$843$\\
Set 2 &$0.15$&$204$&$364$&$5.6$&$136$&$779$\\
\hline
\hline
&$\frac{Y_\pi}{X_\pi}$&$\frac{Y_\rho}{X_\rho}$&$f_\pi$&$f_\rho$&$Z$ \\
&&&(MeV)&& \\
\hline
Exp. &&&$93$&$5.1$&\\
\hline
Set 1 &$-0.842$&$0.169$&$190(93)$&$2.6(5.3)$&$0.70$\\
Set 2 &$-0.847$&$0.169$&$178(89)$&$2.5(5.1)$&$0.71$\\
\hline
\hline
\end{tabular}
\end{table}
Both of pion and rho meson masses agree with the experimental
values well. The current and constituent quark masses are also
reasonable in the ranges of theoretical and experimental
estimations.

The decay constants $f_\pi$ and $f_\rho$ deviate from the
experimental data by a factor 2. This is not very surprising. In
the derivation of the effective Hamiltonian (which we skip in this
work), the current quark field $\Psi$ is renormalized into
constituent quark field. This means that the constituent quark may
include the contributions from configurations other than the
single quark, like $qq\bar{q}$, $qg$ etc as indicated through the
deep inelastic scattering experiments. For example, we may assume
the normalized constituent quark field $\Psi_N$
\begin{equation}
\Psi_N = Z\Psi + \mbox{contributions from $qq\bar{q}$, $qg$, ...}.
\end{equation}
Here we have introduced the factor $Z$. Roughly speaking, $|Z|^2$
is the probability of finding the "pure" current quark field in a
complicated constituent quark. However, from the electro-weak
Lagrangian we know that what really participates in the
electro-weak interactions are the current quarks. In other words,
only the "pure" current quark will contribute to the pion weak
decay constant and rho meson electromagnetic decay constant. In
our calculation we need two real current quarks to annihilate. So
there will be a difference of factor $Z^2$. Decay constants
$f_\pi$ and $f_\rho$ can then be fit to the experimental values if
we set the additional normalization factor $Z\approx 0.7$ (In
Table \ref{table-1} the results are listed in parentheses).

\section{Summary}
The chiral symmetry is studied in an extended constituent quark
potential model which includes the small component effect with RPA
approximation. With some modification of quark interaction in
potential model, $\pi$ meson is still a Goldstone boson. The small
mass of $\pi$ meson can be connected to nonzero current quark mass
from PCAC.  All other mesons such as the vector $\rho$ meson can also
be studied in this model like in the ordinary constituent quark
models. With the small component effect of y-channel contribution, the
well-known KSRF relation of the $\pi$ and $\rho$ meson decay constants
is reestablished in the model.

This extension to constituent quark potential model aims at improving
the constituent quark model to comprise the small component, i.e., the
$y-$ channel quark excitations which are essential to $\pi$ (and K)
mesons. To thoroughly investigate the effect of this extension, we
need perform a full calculation of meson spectroscopy.

\section*{Acknowledgments}
This work is supported by National Natural Science Foundation of
China, Ministry of Education of China, FANEDD and SRF for ROCS,
SEM.


\begin{thebibliography}{10}
\expandafter\ifx\csname url\endcsname\relax
  \def\url#1{\texttt{#1}}\fi
\expandafter\ifx\csname urlprefix\endcsname\relax\def\urlprefix{URL }\fi

\bibitem{Isgur85}
S.~Godfrey, N.~Isgur, Mesons in a relativized quark model with chromodynamics,
  Phys.\ Rev. D32 (1985) 189.

\bibitem{Cotanch00}
F.~J. Llanes-Estrada, S.~R. Cotanch, Phys.\ Rev.\ Lett. 84 (2000) 1102.

\bibitem{Nambu61}
Y.~Nambu, G.~Jona-Lasinio, Phys.\ Rev. 122 (1961) 246.

\bibitem{Bernard84}
V.~Bernard, R.~Brockmann, M.~Schaden, W.~Weise, E.~Werner, Nucl.\ Phys. A412
  (1984) 349.

\bibitem{Klimt90}
S.~Klimt, M.~Lutz, U.~Vogl, W.~Weise, Generalized {$SU(3)$} nambu--jona-lasinio
  model, Nucl.\ Phys. A516 (1990) 429.

\bibitem{Langfeld96}
K.~Langfeld, C.~Kettner, H.~Reinhardt, Nucl.\ Phys. A608 (1996) 331.

\bibitem{DWZ02}
W.~Deng, X.~Chen, D.~LU, L.~Yang, Extended quark potential model from random
  phase approximation, Commun.\ Theor.\ Phys. 38 (2002) 327.

\bibitem{KS1966}
K.~Kawarabayashi, M.~Suzuki, Partially conserved axial-vector current and the
  decays of vector mesons, Phys.\ Rev.\ Lett. 16 (1966) 255.

\bibitem{RF1966}
Riazuddin, Fayyazuddin, Algebra of current componets and decay widths of $\rho$
  and {$K^*$} mesons, Phys.\ Rev. 147 (1966) 1071.

\bibitem{Wilson94}
K.~G. Wilson, T.~S. Walhout, A.~Haringdranath, W.-M. Zhang, R.~J. Perry, S.~D.
  Glazek, Phys.\ Rev. D49 (1994) 6720.

\bibitem{Cotanch99}
D.~Robertson, E.~S. Swanson, A.~P. Szczepaniak, C.-R. Ji, S.~R. Cotanch, Phys.\
  Rev. D59 (1999) 074019.

\bibitem{Cotanch02}
F.~J. Llanes-Estrada, S.~R. Cotanch, Nucl.\ Phys. A697 (2002) 303.

\bibitem{Hooft1974}
G.~'t~Hooft, A planar diagram theory for strong interaction, Nucl.\ Phys. B72
  (1974) 461--473.

\bibitem{Witten79}
E.~Witten, Baryon in the {$1/N$} expansion, Nucl.\ Phys. B160 (1979) 57--115.

\bibitem{DWZ01}
W.~Deng, X.~Chen, D.~Lu, L.~Yang, Meson structures from random phase
  approximation, Phys.\ Lett. B506 (2001) 85--92.

\bibitem{Ring80}
P.~Ring, P.~Schuck, The Nuclear Many-Body Problem, Springer, Berlin, 1980.

\bibitem{Gromes77}
D.~Gromes, Nucl.\ Phys. B131 (1977) 80.

\bibitem{PDG2000}
{Particle Data Group}, Review of particle physics, Euro.\ Phys.\ J. C15 (2000)
  1.

\end{thebibliography}

\end{document}